\documentclass{article}

\usepackage{microtype}
\usepackage{graphicx}
\usepackage{adjustbox} 
\usepackage{booktabs}
\usepackage{hyperref}
\usepackage{xcolor}
\usepackage{fontawesome5}
\usepackage{enumitem}

\newcommand{\hficon}{%
  \IfFileExists{figures/hf-logo.png}{%
    \raisebox{-0.1ex}{\smash{\includegraphics[height=2.2ex]{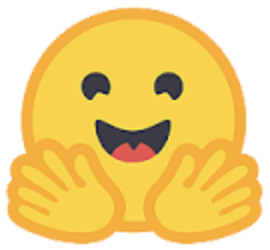}}}%
  }{\textsf{HF}}%
}

\usepackage[accepted]{mlsys2025}

\graphicspath{{figures/}{../figures/}}

\mlsystitlerunning{Lightweight Scheduling for Continuously Batched LLM Serving}

\begin{document}
\twocolumn[\vspace{-5mm}
\mlsystitle{Lightweight Scheduling Policies for Continuously \\Batched LLM Serving Across GPU Architectures}

\begin{mlsysauthorlist}
\begin{tabular}{ccc}
\shortstack{\textbf{Ashman Mehra}\textsuperscript{*}\\
{\small Carnegie Mellon University,}\\
{\small School of Computer Science}\\
{\ttfamily ashmanm@andrew.cmu.edu}}
&
\shortstack{\textbf{Akshan Agrawal}\textsuperscript{*}\\
{\small Carnegie Mellon University,}\\
{\small School of Computer Science}\\
{\ttfamily akshana@andrew.cmu.edu}}
&
\shortstack{\textbf{Aryeh Carmi}\textsuperscript{*}\\
{\small Carnegie Mellon University,}\\
{\small School of Computer Science}\\
{\ttfamily acarmi@andrew.cmu.edu}}
\end{tabular}
\par\vspace{1mm}
{\small \textsuperscript{*}Equal contribution}
\end{mlsysauthorlist}

\mlsyskeywords{Machine Learning Systems, LLM Serving, Scheduling, Continuous Batching}

\vskip 0.1in

\begin{abstract}
Large language model serving is increasingly limited not only by model size and GPU speed, but also by scheduling during autoregressive decoding, where continuous batching interleaves requests at token granularity as sequences arrive and finish. We study whether lightweight scheduling policies can improve latency-throughput tradeoffs without changing model architectures, kernels, or deploying learned schedulers. We implement a Python/PyTorch continuous-batching testbench with real Hugging Face execution and evaluate eight policies across 1,152 GPU-model-scheduler-workload runs using ShareGPT-derived traffic, six Qwen, Gemma, and Mistral models, and Modal L40S, A100, and H100 GPUs. Scheduler choice changes throughput, TTFT, completion latency, and tail behavior under fixed model, GPU, workload, batch size, and prefill budget. Aging shortest-prompt-first is the strongest practical baseline in many L40S and mixed-workload settings, while oracle remaining-length scheduling shows that job-size knowledge alone is insufficient because batching and hardware effects also shape performance.
The custom dataset has been uploaded to Hugging Face: \href{https://huggingface.co/datasets/ashman1705/LLMScheduling}{\textcolor{black}{\hficon}\ \texttt{\textbf{dataset}}}.
\end{abstract}
]

\section{Introduction}

Serving large language models is fundamentally different from serving traditional neural networks because generation is autoregressive. A request is not completed by a single forward pass; instead, the system repeatedly executes decode steps, each producing one new token while carrying forward a growing key-value cache. In an online setting, many requests with different prompt lengths, output lengths, and arrival times compete for the same GPU. This makes scheduling a first-order systems problem: the runtime must decide which requests receive service at each decode step while keeping the GPU sufficiently full to maintain high throughput.

Continuous batching has become a standard technique for improving LLM serving efficiency. Rather than waiting for a fixed batch of requests to finish together, the serving system admits new requests as old requests complete. This improves utilization and reduces idle time under realistic online traffic. However, continuous batching does not eliminate head-of-line blocking. If long generations remain in the active batch for many decode iterations, they can delay shorter jobs that would otherwise finish quickly. A naive first-in, first-out policy is simple and fair by arrival order, but it can be inefficient when request lengths vary widely.

This project investigates a practical question: how much can continuously batched LLM serving improve using only lightweight scheduling changes? Rather than proposing a new model architecture, a new inference kernel, or a full production serving stack, we isolate the scheduling policy as the variable of interest. We compare simple policies that use request metadata such as prompt length, target or estimated remaining decode work, service tier, waiting time, and observed completion statistics. The goal is not to find a universally optimal scheduler, but to understand how scheduler choice affects latency, throughput, and tail behavior across workloads, models, and GPUs.

This research makes three contributions: 
\begin{enumerate}[leftmargin=*, nosep]
    \item We implement a pluggable continuous-batching testbench for comparing LLM serving schedulers under fixed runtime constraints. The testbench executes real Hugging Face causal language models using PyTorch, batched prefill, cached one-token decode, and per-request metric logging.
    \item We construct a ShareGPT-derived workload suite with bursty, low-load, steady, uniform, and mixed short/long request patterns, allowing scheduler behavior to be evaluated under both benign and head-of-line-blocking-heavy traffic. The custom dataset includes 22 categories of tests, each with 8 test cases (100+ requests). The dataset is also made open-source for benchmarking.
    \item We evaluate eight scheduling policies across six open-weight models and three GPU classes, measuring throughput, time-to-first-token, completion latency, p95/p99 latency, batch efficiency, starvation behavior.
\end{enumerate}
Our results show that scheduling policy is a meaningful systems lever: aging shortest-prompt-first is a strong practical baseline, while oracle remaining-length scheduling shows that job-size knowledge alone does not determine performance because batching and hardware effects also shape the latency-throughput tradeoff.

\section{Problem Definition and Research Questions}

\subsection{Continuous-Batching Decode Scheduling}

In autoregressive LLM inference, each request consists of an initial prefill stage followed by a decode stage. During prefill, the model processes the prompt and initializes the request's KV cache. During decode, the model repeatedly generates one token conditioned on the previous tokens. Since decode proceeds incrementally, the serving runtime has repeated opportunities to choose which requests to include in the next GPU step.

The runtime maintains future requests that have not yet arrived, active requests that are eligible for service, and completed requests. When the wall-clock replay time reaches a request's arrival time, the request becomes active. At each decode step, the scheduler ranks active requests and selects at most \path{max_batch_size} requests for service. In our experiments, \path{max_batch_size} is fixed at 16. For requests receiving their first service, the testbench performs batched prefill; for requests with an existing KV cache, it performs cached one-token decode.

The scheduler affects at least four outcomes. It affects throughput because larger or more stable batches improve GPU utilization. It affects time-to-first-token because admission and first service determine how quickly a user sees an initial response. It affects completion latency because scheduling determines how long each request takes from arrival to final token. It also affects tail latency and fairness because aggressive short-job policies can improve average latency while delaying long or low-priority requests.

The core challenge is that a deployable scheduler usually does not know the true remaining work of a request. Prompt length is visible at arrival time, but output length is often unknown. In our trace-driven testbench, target output length is known from the ShareGPT response length and is used to determine when a request completes. This lets us compare both practical heuristics and an oracle-style shortest-remaining-tokens policy that uses the trace-derived remaining output length.

\subsection{Research Questions}

\paragraph{RQ1: Do short-job-aware schedulers improve latency and throughput compared with FIFO continuous batching?}
FIFO is attractive because it is simple and predictable, but it can allow long requests to occupy decode capacity while shorter requests wait. Policies inspired by shortest-job-first or shortest-remaining-processing-time may reduce mean and tail latency by finishing short requests earlier.

\paragraph{RQ2: How much does estimation quality matter?}
Some policies use visible prompt length as a proxy for job size, while others estimate remaining decode length from prompt length, request class, and online observations. We evaluate whether these estimates improve performance and whether inaccurate estimates can hurt compared with simpler heuristics.

\paragraph{RQ3: Does the best scheduling policy depend on hardware, model, and workload?}
GPU architectures differ in memory bandwidth, compute throughput, and batching behavior. Models differ in size, context handling, and decode cost. Workloads differ in arrival pattern and length distribution. We therefore expect no single scheduler to dominate universally.

\section{Related Work}

Orca introduced iteration-level scheduling for transformer serving, showing that generative inference can benefit from scheduling at the granularity of decode iterations rather than entire requests~\cite{yu2022orca}. This idea is central to modern continuous batching: requests can enter and leave the batch dynamically, allowing the system to maintain utilization under variable-length generation. Our work builds on this observation, but focuses on comparing lightweight policies inside a controlled continuous-batching testbench rather than designing a new distributed serving system.

FastServe further shows that preemption and scheduling are central to interactive LLM serving, using token-level preemption and a skip-join multi-level feedback queue scheduler to reduce head-of-line blocking~\cite{wu2023fastserve}. Our study differs by holding the serving mechanism fixed and comparing lightweight scheduler choices across models, GPUs, and workload families.

vLLM and PagedAttention showed that KV-cache memory management is critical for high-throughput LLM serving~\cite{kwon2023pagedattention}. By reducing memory fragmentation and enabling efficient paging of KV-cache blocks, such systems increase the number of concurrent sequences that can be served. Our project is complementary and builds on the same core insight: once a continuous-batching abstraction makes many concurrent sequences feasible, scheduler choice becomes a practical lever for shaping latency-throughput tradeoffs. Rather than replacing systems like vLLM, we study which active requests should receive decode service under continuous batching, providing actionable guidance that can be layered on top of strong execution engines.

Recent serving systems also show that the interaction between prefill, decode, and batching policy is a first-order performance issue. SARATHI uses chunked prefills and decode-maximal batching to reduce prefill/decode imbalance~\cite{agrawal2023sarathi}, while DistServe disaggregates prefill and decode resources to optimize TTFT and TPOT service objectives~\cite{zhong2024distserve}. These systems motivate our emphasis on scheduler behavior, although our evaluation intentionally keeps the serving architecture fixed.

Classical scheduling theory suggests that shortest-job-first and shortest-remaining-processing-time can reduce mean completion time when job sizes are known or well estimated; Schrage's  (Shortest Remaining Processing Time) result is a canonical reference for this framing~\cite{schrage1968srpt}. LLM serving complicates this because request cost depends on both prompt length and generated output length. Recent learning-based work, including learning-to-rank approaches for LLM scheduling, predicts output length or relative job cost to approximate these policies~\cite{fu2024efficient}. Our project explores a lower-overhead design space: heuristic policies based on prompt length, trace-derived remaining length, online output-length estimates, aging, and service-tier fairness.

\section{System Overview}

\begin{figure*}[t]
\centering
\includegraphics[width=0.8\textwidth]{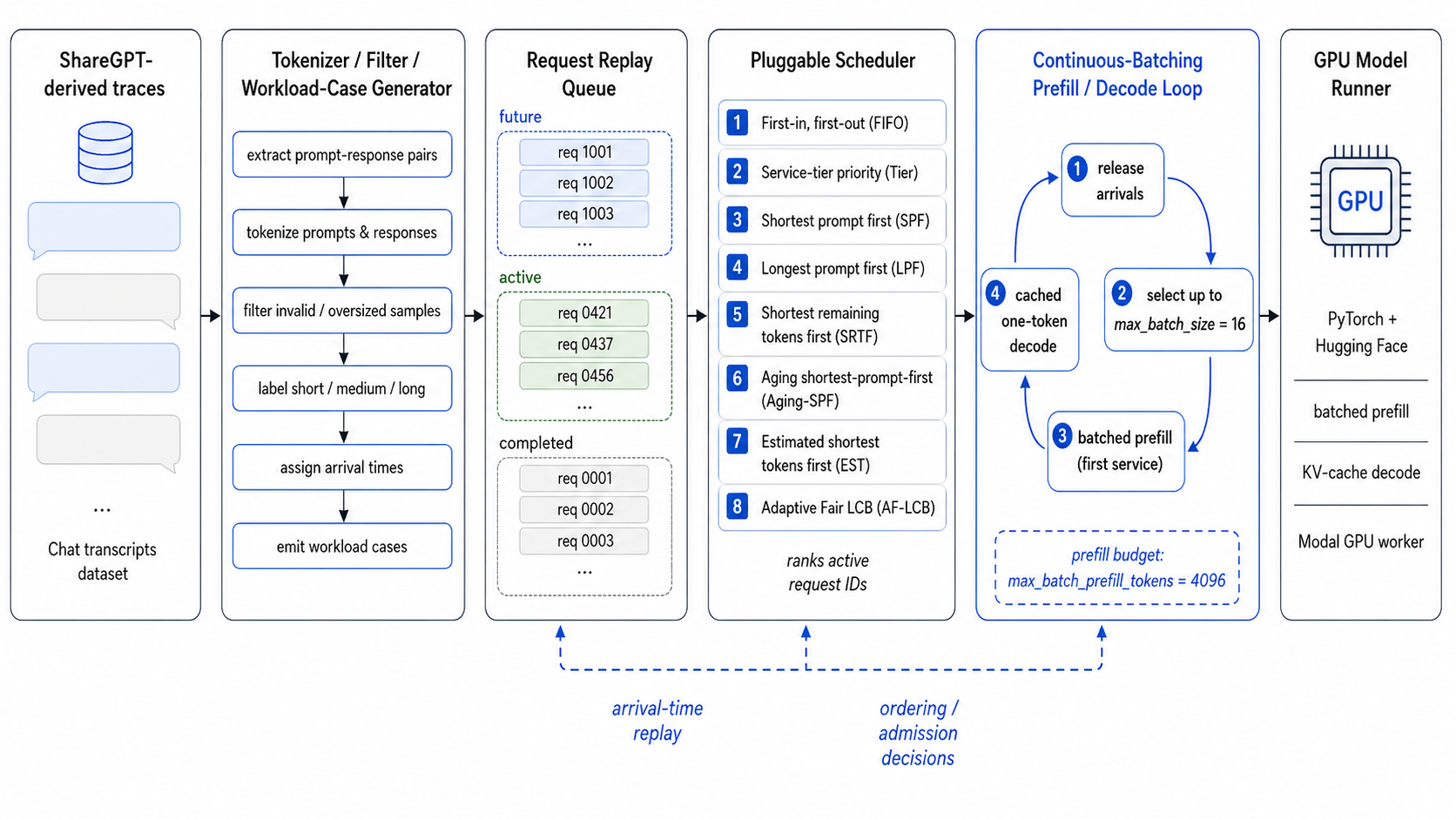}
\vspace{-10mm}
\caption{System overview of the replay-based continuous-batching LLM serving testbench.}
\label{fig:system-overview}
\end{figure*}

Figure~\ref{fig:system-overview} shows the architecture of our evaluation system. The system is a replay-based continuous-batching LLM serving testbench. It takes ShareGPT-derived workload cases, replays request arrivals according to the case definition, ranks active requests with a pluggable scheduler, executes the selected batch on a GPU-backed Hugging Face model runner, and logs per-request and per-step metrics.

The system has four main components. The workload generator builds scheduler-ready cases from ShareGPT-style conversations. It extracts human-to-assistant request/response pairs, tokenizes prompts and responses, filters invalid or oversized samples, labels requests as short, medium, or long, assigns arrival times, and emits workload cases. The scheduler receives snapshots of waiting, active, and completed requests and returns an ordering of active request IDs. The model execution layer uses PyTorch and Hugging Face Transformers to run batched prefill and cached one-token decode. The metrics logger records request-level statistics such as queue delay, TTFT, completion latency, generated tokens, scheduled steps, decode time, service tier, and request class, along with aggregate throughput and batch efficiency.

The same workload cases are replayed for every scheduler under fixed controls. The maximum decode batch size is 16. The prefill-token budget is 4096 padded tokens per prefill sub-batch, implemented as max prompt length in the sub-batch times the number of prefilling requests. These controls ensure that performance differences primarily reflect scheduler behavior rather than changes in workload composition or batch constraints.

\section{Methodology}

\subsection{Trace-Driven Workload Construction}

We evaluate on ShareGPT-derived conversational workloads~\cite{sharegpt}. The workload pipeline builds cases from the \path{Aeala/ShareGPT_Vicuna_unfiltered} ShareGPT-style dataset using the Hugging Face datasets library. Prompts and target responses are tokenized with the Hugging Face GPT-2 tokenizer for workload construction. Samples with fewer than four input or output tokens are removed, and prompts or responses above the 4096-token source cap are filtered out. The source pool is deduplicated and trimmed to 50,000 cleaned request records.

Each request stores the prompt, input-token count, target output-token count, arrival time, priority class, service tier, and source metadata. Priority class is assigned from prompt length: short requests have fewer than 128 input tokens, long requests have more than 1024 input tokens, and medium requests fall between those thresholds. The target output-token count comes from the ShareGPT response length. During model execution, this target length determines the default stopping point, subject to model context limits and the configured per-request generation cap.

The workload generator can emit eight ShareGPT-derived workload categories and 22 seeded cases per category, for 176 total generated cases. The final reported analysis uses one representative seeded case from each category, so the reported run matrix is at the GPU-model-scheduler-category level rather than an average over all 22 seeds. The categories stress different scheduling behaviors: steady short traffic, steady mixed traffic, bursty conversational traffic, bursty long-request traffic, flash-crowd spikes, low-load traffic, short/long mixes, and a uniform baseline. Poisson arrivals are generated with exponential interarrival times, while bursty workloads use clustered arrivals with random intra-cluster spacing. This gives a mix of average-case, low-contention, and head-of-line-blocking-heavy workloads.

\subsection{Scheduler Policies}

All schedulers implement the same interface: at each decode step, they rank active requests and the runtime serves the first \path{max_batch_size} requests in that ordering. Table~\ref{tab:scheduler-policies} summarizes each priority signal. Unless otherwise stated, ties are broken by \path{arrival_time} and then \path{request_id}.

\begin{table}[t]
\caption{Scheduler policies and priority signals. Policy abbreviations used in the results appear in parentheses.}
\label{tab:scheduler-policies}
\vskip 0.15in
\begin{center}
{\scriptsize
\setlength{\tabcolsep}{4pt}
\renewcommand{\arraystretch}{1.08}
\begin{tabular}{@{}p{0.48\linewidth}p{0.45\linewidth}@{}}
\toprule
POLICY & PRIORITY SIGNAL \\
\midrule
First-in, first-out (FIFO) & Arrival order. \\
Service-tier priority (Tier) & Service tier, then arrival order. \\
Shortest prompt first (SPF) & Visible prompt length. \\
Longest prompt first (LPF) & Negative prompt length. \\
Shortest remaining tokens first (SRTF) & GPT-3 output length. \\
Aging shortest-prompt-first (Aging) & Prompt waiting age. \\
Estimated shortest tokens first (ESTF) & Online estimated output length. \\
Adaptive fair LCB (AF-LCB) & Estimated remaining work, KV pressure, age bonus, and tier deficit. \\
\bottomrule
\end{tabular}
}
\end{center}
\end{table}

All policies except shortest remaining tokens first are deployable from request metadata and online observations. Shortest remaining tokens first is oracle-style because it uses the trace-derived target output length. Estimated shortest tokens first updates per-class and global output-length averages from completed requests. Adaptive fair LCB combines estimated remaining work, uncertainty, starvation prevention, service-tier fairness, and KV-cache pressure, intentionally trading off short-job preference against starvation and service-tier debt.

\subsection{Continuous-Batching Testbench Implementation}

The testbench is implemented in Python 3.11 using PyTorch and Hugging Face Transformers on Modal GPU workers. The model runner loads each model with Hugging Face APIs, performs batched prefill with cached KV state for newly admitted requests, and then runs one-token greedy decode for the selected active batch at each scheduling step.

The execution is hybrid: arrivals and target output lengths are trace-driven, but model execution is real. Requests stop after reaching the trace-derived target output length, capped by model context and a 1024 generated-token limit, so every scheduler receives the same intended work.

The runtime owns request state, arrival release, prefill splitting, decode execution, timing, and metric logging. The scheduler only sees request metadata and runtime progress through SchedulerSnapshot objects and returns an ordering of active request IDs, which makes each policy comparable under identical model, case, batch, and prefill settings.

\subsection{Adopted Components and Implemented Components}

To avoid conflating infrastructure with our contribution, we separate reused components from the artifacts we implemented for the project.

We adopted the ShareGPT-style conversational data source, Hugging Face tokenizers and model-loading APIs, PyTorch tensor execution, Modal GPU infrastructure, and open-weight Qwen, Gemma, and Mistral models. We also adopt the general continuous-batching abstraction used by modern LLM serving systems as the conceptual setting.

We implemented the workload curation pipeline, request-class labeling, arrival-process generation, scheduler API, eight scheduler policies, continuous-batching testbench, Hugging Face prefill/decode runner, Modal experiment runner, and metrics collection. The repository also contains optional Triton and \path{torch.compile} kernel-proxy code, but kernel optimization was not part of the final scheduler evaluation. The final project scope is an empirical scheduler study, not a kernel optimization paper.

\section{Experimental Setup}

\subsection{Models and Hardware}

We evaluate six open-weight Hugging Face model repositories from the Qwen, Gemma, and Mistral families: \path{Qwen/Qwen3-0.6B}, \path{Qwen/Qwen3.5-2B}, \path{Qwen/Qwen2.5-1.5B-Instruct}, \path{google/gemma-4-E2B-it}, \path{google/gemma-4-E4B-it}, and \path{mistralai/Ministral-3-3B-Instruct-2512}.

Experiments are run on Modal GPU workers across NVIDIA L40S, A100, and H100 GPUs. This allows us to test whether scheduler behavior is stable across GPU architectures or whether hardware changes the preferred policy.

\subsection{Experimental Controls}

For each GPU and model setting, every scheduler is evaluated on the same representative workload cases. We fix \path{max_batch_size} at 16 and \path{max_batch_prefill_tokens} at 4096. The final analysis contains 1,152 executed GPU-model-scheduler-category runs: three GPUs, six models, eight schedulers, and eight workload categories. Each reported category run uses the same selected workload case for all schedulers in that GPU-model setting; it is not an aggregate over the 22 available seeds. The model, request arrivals, target output lengths, and runtime limits are held constant while the scheduler changes. This control is important because otherwise a scheduler might appear better simply because it saw easier traffic or admitted larger batches.

\subsection{Metrics}

Throughput is measured as generated tokens per second over wall-clock runtime, with completed requests tracked per run. Time-to-first-token measures arrival-to-first-token responsiveness, mean completion latency measures arrival-to-final-token time, and p95/p99 latency captures tail user experience, especially in mixed workloads where mean latency can improve while long or unlucky requests worsen. Batch efficiency uses average batch size, decode-step count, and average decode-step time to show whether a scheduler improves queueing behavior, GPU utilization, or both. Fairness and starvation behavior are evaluated through request-length breakdowns, service-tier behavior, and mixed short/long workloads.

\section{Results and Analysis}

We organize the results around the three research questions. First, scheduler choice changes the latency-throughput frontier: runs with the same model, GPU, workload family, batch limit, and prefill budget land at different points depending only on the scheduler. Second, short-job-aware policies are often beneficial, but the strongest versions include aging or fairness mechanisms to avoid starving longer requests. Third, scheduler behavior is hardware-dependent; policies that are competitive on one GPU are not necessarily best on another.

\subsection{Scheduler Choice Shifts the Latency-Throughput Frontier}

\begin{figure*}[t]
\centering
\includegraphics[width=0.98\textwidth]{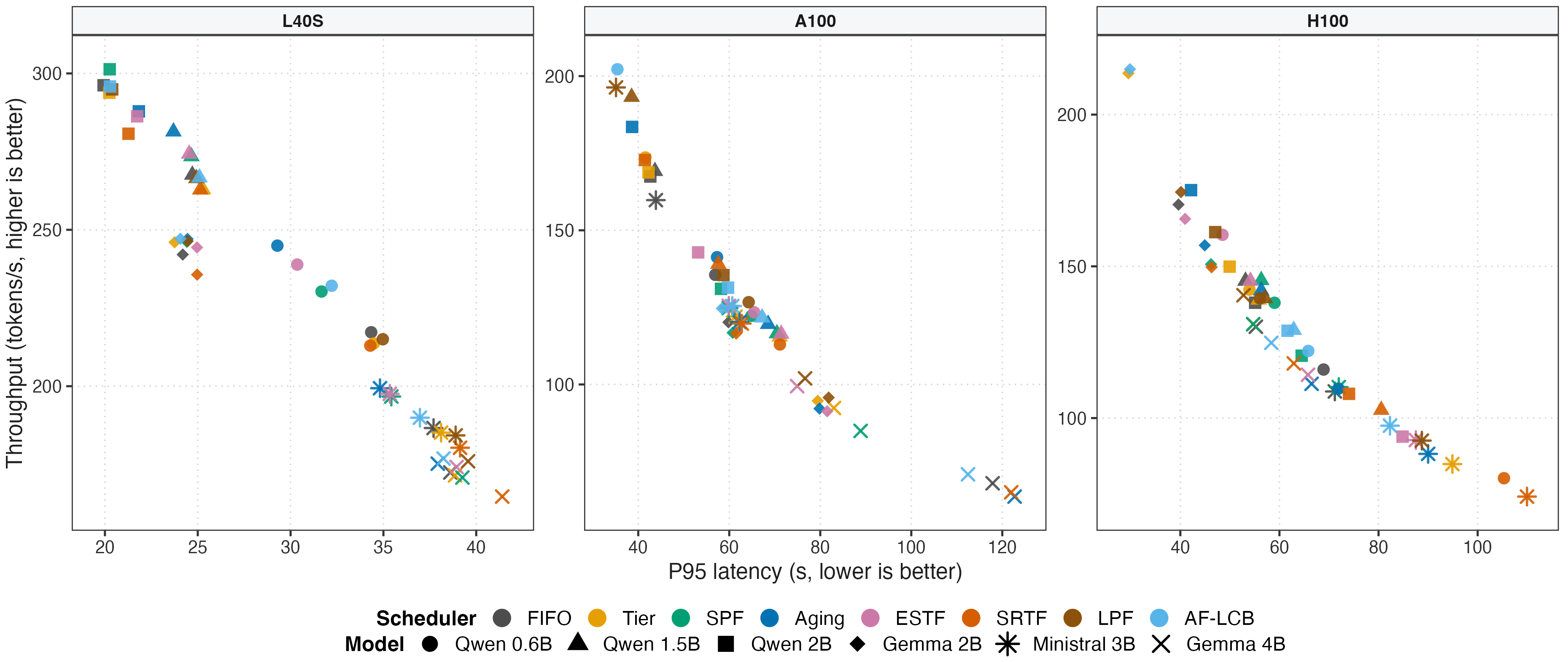}
\caption{Latency/throughput frontier across GPU, model, and scheduler runs, faceted by GPU with independent axis ranges. Each point is one GPU-model-scheduler average across workload categories; color encodes scheduler and marker shape encodes model. Lower p95 latency and higher throughput are preferred. SPF denotes shortest prompt first, ESTF estimated shortest tokens first, SRTF shortest remaining tokens first, LPF longest prompt first, and AF-LCB adaptive fair LCB.}
\label{fig:latency-throughput}
\end{figure*}

\begin{figure*}[t]
\centering
\includegraphics[width=0.98\textwidth]{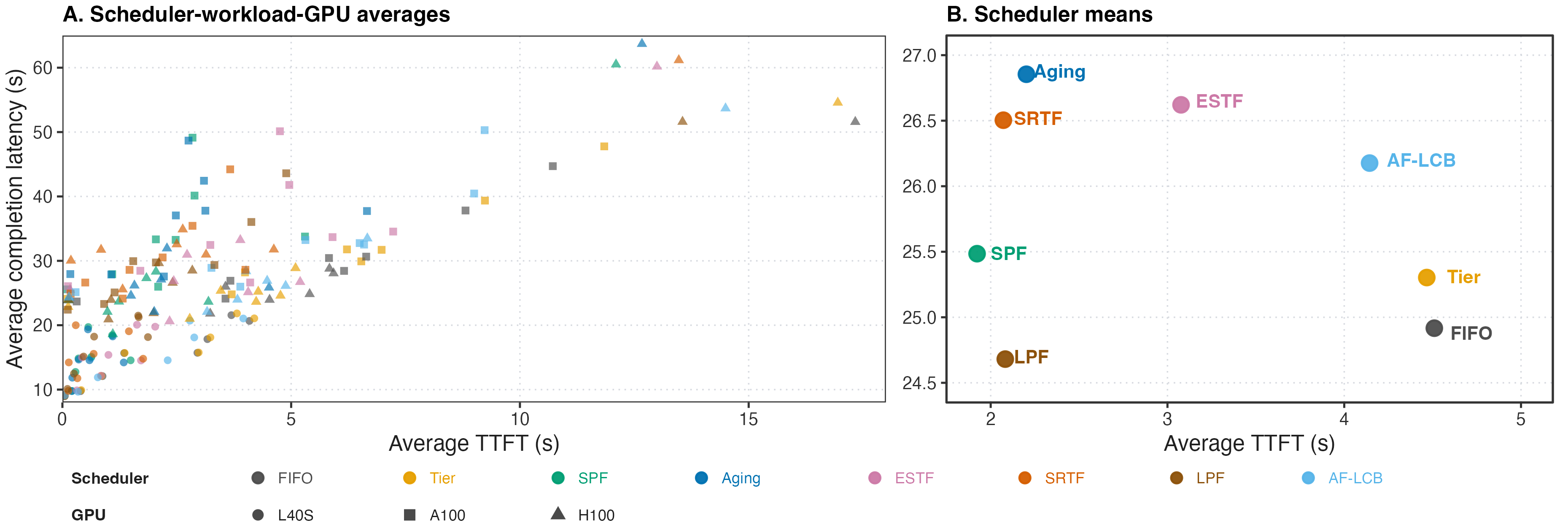}
\caption{TTFT vs. completion latency. Left: each point is a scheduler-workload-GPU average across models; circles, squares, and triangles denote L40S, A100, and H100. Right: labeled scheduler means across all runs. Lower-left points give better interactivity and completion time.}
\label{fig:ttft-completion}
\end{figure*}

Figure~\ref{fig:latency-throughput} shows the latency-throughput frontier across GPU, model, and scheduler runs. Each point averages across workload categories for one GPU-model-scheduler configuration. Points with lower p95 latency and higher throughput are preferred. Faceting by GPU with independent axis ranges makes within-GPU model and scheduler variation easier to read, while the model markers preserve the model-level variation that drives much of the frontier. Within these model/GPU envelopes, scheduler color still shifts points toward or away from the efficient region, indicating that scheduling policy is a meaningful performance knob even without changing the model or serving stack.

Model size and GPU type strongly affect absolute performance. Smaller models and faster configurations tend to occupy the high-throughput, lower-latency region, while larger or slower configurations move toward lower throughput and higher p95 latency. The scheduler effect is smaller than the model/hardware effect, but it is still visible within each region. This is the main empirical pattern: scheduling is useful, but it should be tuned against the target model, hardware, workload mix, and metric.

Figure~\ref{fig:ttft-completion} shows why TTFT and completion latency must be reported together. The left panel shows that scheduler-workload-GPU combinations spread substantially in both dimensions, especially under bursty traffic. The right panel summarizes this variation with scheduler means: short-job-aware policies such as shortest prompt first and aging SPF move left by reducing average TTFT, but they do not always minimize completion latency. FIFO, service-tier priority, and longest-prompt-first can have slower first-token response while remaining competitive on completion latency. A scheduler evaluation that reports only one of these metrics would miss this tradeoff.

\begin{figure*}[t]
\centering
\includegraphics[width=0.98\textwidth]{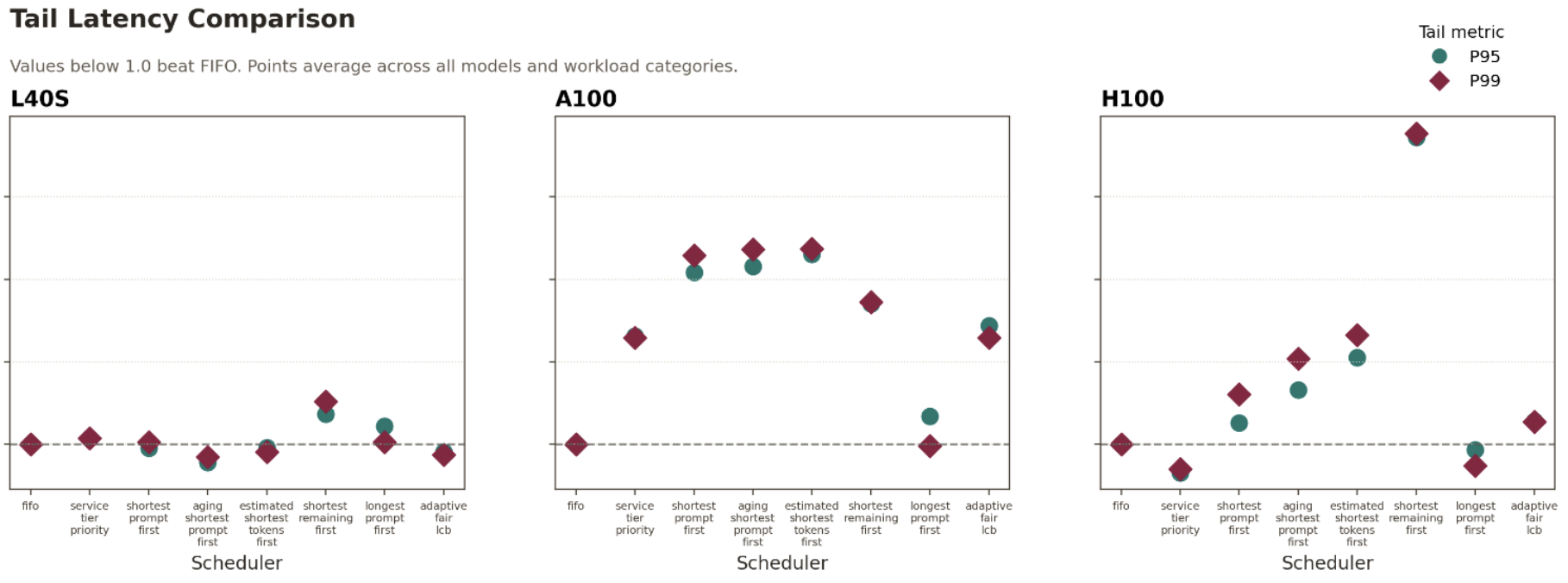}
\caption{Tail-latency comparison across GPUs, with p95 and p99 normalized to FIFO. Values below 1.0 improve on FIFO.}
\label{fig:tail-latency}
\end{figure*}

\subsection{Aging-Based Short-Job Scheduling Is a Strong L40S Baseline}

\begin{table*}[t]
\caption{L40S throughput leaderboard across schedulers. Mean throughput in generated tokens/s for each model-scheduler pair on NVIDIA L40S, averaged across workload categories. Higher is better.}
\label{tab:l40s-throughput}
\vskip 0.15in
\begin{center}
{\scriptsize
\setlength{\tabcolsep}{2.4pt}
\renewcommand{\arraystretch}{1.04}
\begin{adjustbox}{max width=0.95\textwidth}
\begin{tabular}{lcccccccc}
\toprule
Model & FIFO & \shortstack{Service-tier\\priority (Tier)} & \shortstack{Shortest prompt\\first (SPF)} & \shortstack{Aging shortest-\\prompt-first (Aging)} & \shortstack{Estimated shortest\\tokens first (ESTF)} & \shortstack{Shortest remaining\\tokens first (SRTF)} & \shortstack{Longest prompt\\first (LPF)} & \shortstack{Adaptive fair\\LCB (AF-LCB)} \\
\midrule
Qwen3-0.6B & 217.2 & 213.7 & 230.3 & \textbf{244.9} & 238.9 & 213.0 & 215.0 & 232.1 \\
Qwen2.5-1.5B-Instruct & 267.6 & 262.9 & 273.5 & \textbf{281.4} & 274.3 & 262.8 & 266.5 & 266.7 \\
Qwen3.5-2B & 296.2 & 293.8 & \textbf{301.3} & 287.9 & 286.3 & 280.7 & 294.9 & 295.8 \\
gemma-4-E2B-it & 242.1 & 246.0 & 246.1 & \textbf{247.1} & 244.4 & 235.7 & 246.4 & \textbf{247.1} \\
Ministral-3-3B-Instruct-2512 & 186.6 & 185.1 & 196.7 & \textbf{199.4} & 197.6 & 180.3 & 184.3 & 189.9 \\
gemma-4-E4B-it & 172.4 & 171.5 & 170.8 & 175.2 & 174.2 & 164.7 & 176.0 & \textbf{176.9} \\
\bottomrule
\end{tabular}
\end{adjustbox}
}
\end{center}
\end{table*}

We present L40S tables as a detailed representative breakdown; cross-GPU trends are summarized in Figures~\ref{fig:latency-throughput} and~\ref{fig:tail-latency}. Tables~\ref{tab:l40s-throughput} and~\ref{tab:l40s-tail-latency} quantify RQ1 on L40S. Every model has a non-FIFO scheduler with higher throughput, with gains from 1.7\% for Qwen3.5-2B to 12.8\% for Qwen3-0.6B. The best p95 scheduler improves tail latency for five of six models, with gains up to 14.7\%; the exception is Qwen3.5-2B, where FIFO's 19.93 second mean p95 is best. Aging shortest-prompt-first is the best or tied-best throughput policy for four models and the best p95 policy for four models. These results answer RQ1 with a qualified yes: short-job-aware scheduling often helps, but aging is important because it gives long-waiting requests increasing priority and reduces starvation risk.

\subsection{Better Estimates Do Not Always Produce Better Schedules}

RQ2 asked whether estimation quality matters. Tables~\ref{tab:l40s-throughput} and~\ref{tab:l40s-tail-latency} show that it does, but more complex estimates do not automatically dominate simpler heuristics. Prompt length is an imperfect proxy for total request cost, yet shortest prompt first and aging SPF perform well in several settings. Estimated shortest tokens first and adaptive fair LCB are competitive but not universally best. Most importantly, shortest remaining tokens first has oracle-style access to trace-derived remaining output length, yet it often fails to dominate and can hurt tail latency. The takeaway is that scalar job-size minimization is not the whole objective in GPU serving: batch stability, prefill behavior, cache pressure, aging, and fairness also matter.

\subsection{Scheduler Behavior Depends on GPU Architecture}

RQ3 is supported by the cross-GPU results. Figure~\ref{fig:latency-throughput} separates the latency-throughput frontier by GPU and shows that scheduler behavior changes with hardware and model. L40S places the smaller-model runs in a high-throughput, lower-latency region, while A100 and H100 runs occupy different parts of the frontier and show different scheduler ordering. This means scheduler choice should not be treated as a fixed algorithmic fact independent of the deployment target.

Figure~\ref{fig:tail-latency} shows p95 and p99 tail latency normalized to FIFO on each GPU. The ranking changes by architecture: L40S favors aging shortest-prompt-first on p95, A100 is less friendly to short-job policies, and H100 gives stronger relative results to service-tier and longest-prompt-first policies. P95 and p99 also do not move together for every scheduler. Shortest remaining first is the clearest warning case: even with oracle-style remaining-length information, it can increase tail latency when the policy interacts badly with batching and hardware effects.

\begin{figure}[t]
\centering
\includegraphics[width=0.94\columnwidth]{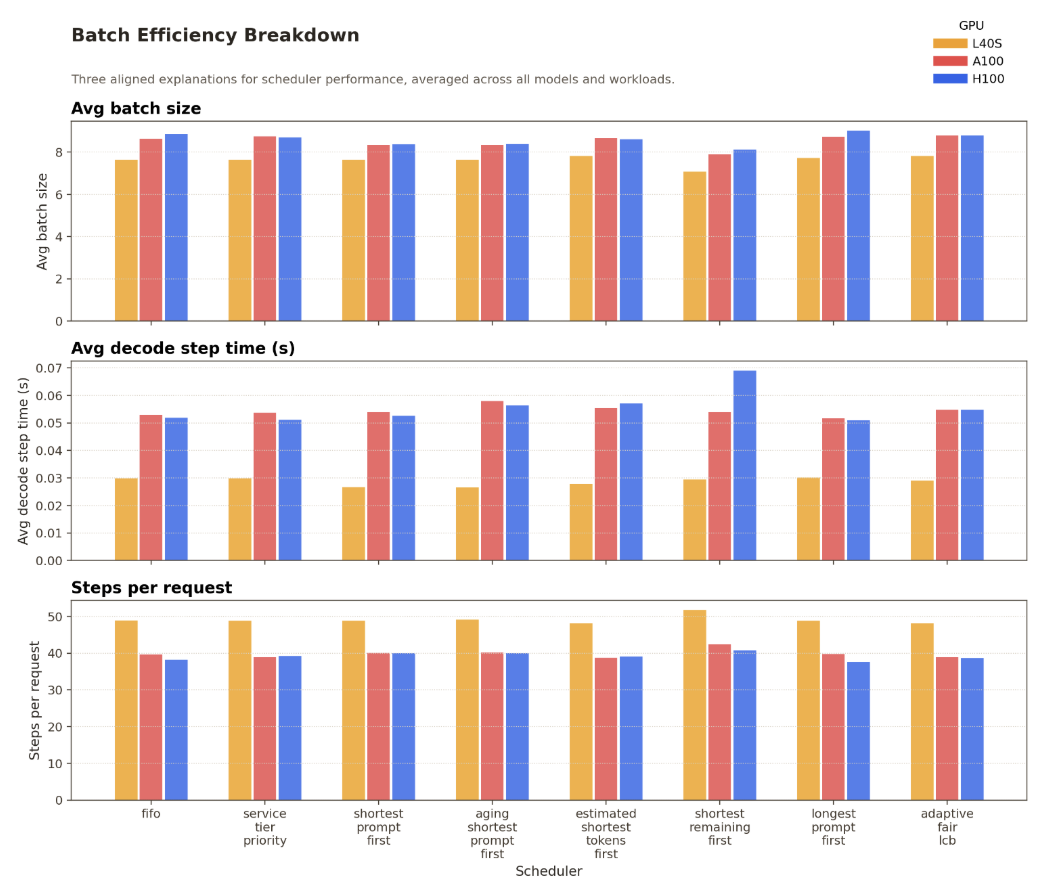}
\caption{Batch efficiency breakdown across schedulers and GPUs: average batch size, decode-step time, and steps per request.}
\label{fig:batch-efficiency}
\end{figure}

Figure 5 shows that LLM serving efficiency depends on the interplay between GPU architecture and scheduling, with the L40S consistently achieving faster decode times with smaller batches than the A100 and H100. Because schedulers like “Shortest Remaining First” can trigger latency anomalies on high-end hardware like the H100, there is no universal “best” configuration. Consequently, the scheduler should be treated as a tunable deployment-time variable to balance hardware-specific memory constraints with throughput and latency targets.


\begin{table}[t]
\caption{L40S p95 latency in seconds, averaged across workload categories. Lower is better.}
\label{tab:l40s-tail-latency}
\vskip 0.15in
\centering
{\footnotesize
\setlength{\tabcolsep}{2.0pt}
\renewcommand{\arraystretch}{1.04}
\begin{adjustbox}{max width=\columnwidth}
\begin{tabular}{lrrrrrrrr}
\toprule
Model & FIFO & Tier & SPF & Aging & ESTF & SRTF & LPF & AF-LCB \\
\midrule
Qwen 0.6B & 34.34 & 34.48 & 31.67 & \textbf{29.29} & 30.36 & 34.29 & 34.97 & 32.21 \\
Qwen 1.5B & 24.71 & 25.28 & 24.66 & \textbf{23.69} & 24.54 & 25.14 & 24.93 & 25.09 \\
Qwen 2B & \textbf{19.93} & 20.24 & 20.26 & 21.83 & 21.74 & 21.27 & 20.39 & 20.27 \\
Gemma 2B & 24.19 & \textbf{23.75} & 24.41 & 24.45 & 24.96 & 24.97 & 24.46 & 24.06 \\
Min. 3B & 37.70 & 38.11 & 35.43 & \textbf{34.82} & 35.35 & 39.13 & 38.89 & 36.97 \\
Gemma 4B & 38.61 & 38.86 & 39.25 & \textbf{37.93} & 38.94 & 41.40 & 39.56 & 38.24 \\
\bottomrule
\end{tabular}
\end{adjustbox}
}
\end{table}

\section{Discussion and Limitations}

This project shows that lightweight scheduling can improve LLM serving behavior, but the evaluation is still a controlled testbench rather than a full production serving stack. Production systems may add admission control, KV-cache eviction, cancellation, speculative decoding, streaming overheads, and network variability. The ShareGPT-derived workload suite also may not match code generation, long-document summarization, or agentic traffic. Because our workloads are replay-based, they do not model feedback effects such as users cancelling slow requests or traffic shifts caused by streaming responsiveness.

The testbench is trace-driven with respect to output length. This keeps the target work fixed across schedulers, but gives shortest remaining tokens first more information than a real online scheduler would normally have. We measure serving latency and throughput, not semantic output quality, and kernel optimization is not part of the evaluated contribution. The practical deployment implication is to treat the scheduler as a low-cost tuning parameter rather than a universal default: FIFO and service-tier priority remain useful for predictability, aging SPF is strong for interactive latency under mixed workloads, and estimated or adaptive policies should be validated on target traffic.

\paragraph{Key takeaways.}
RQ1: short-job-aware scheduling often improves latency and sometimes throughput, especially when aging prevents long-request starvation. RQ2: better length estimates help only when they preserve batching stability and fairness; oracle remaining length is not enough by itself. RQ3: the best scheduler changes with GPU, model, and workload, so serving systems should expose scheduling as a deployment-time control.

\section{Conclusion}

This project evaluates lightweight scheduling policies for continuously batched LLM serving across workloads, models, and GPUs. Using a testbench with real batched prefill and KV-cache decode, we compare eight schedulers on ShareGPT-derived workloads. The key result: scheduler choice materially affects throughput and tail latency even under fixed system settings. Aging shortest-prompt-first is a strong practical baseline, especially on L40S and mixed workloads, but the best choice is deployment-dependent. Oracle remaining-length results show job-size knowledge alone is insufficient—batching and hardware effects also matter—so scheduling should be treated as a workload-aware deployment parameter.

\bibliography{references}
\bibliographystyle{mlsys2025}

\end{document}